\documentclass[sigconf]{acmart}

\usepackage{booktabs}

\usepackage{graphicx}
\usepackage{comment}
\usepackage{caption}
\usepackage{subcaption}
\usepackage{enumitem}
\usepackage{tablefootnote}
\begin{document}

\title{On Application of Learning to Rank for E-Commerce Search}

\author{Shubhra Kanti Karmaker Santu}
\affiliation{University of Illinois\\ Urbana-Champaign (UIUC)}
\email{karmake2@illinois.edu}

\author {Parikshit Sondhi}
\affiliation{WalmartLabs}
\email{psondhi@walmartlabs.com}
       
\author{ChengXiang Zhai}
\affiliation{University of Illinois\\ Urbana-Champaign (UIUC)}
\email{czhai@illinois.edu}

\begin{abstract}
E-Commerce (E-Com) search is an emerging important new application of information retrieval. Learning to Rank (LETOR) is a general effective strategy for optimizing search engines, and is thus also a key technology for E-Com search. While the use of LETOR for web search has been well studied, its use for E-Com search has not yet been well explored. In this paper, we discuss the practical challenges in applying learning to rank methods to E-Com search, including the  challenges in feature representation, obtaining reliable relevance judgments, and optimally exploiting multiple user feedback signals such as click rates, add-to-cart ratios, order rates, and revenue. We study these new challenges using experiments on industry data sets and report several interesting findings that can provide guidance on how to optimally apply LETOR to E-Com search:  First, popularity-based features defined solely on product items are very useful and LETOR methods were able to effectively optimize their combination with relevance-based features. Second, query attribute sparsity raises challenges for LETOR, and selecting features to reduce/avoid sparsity is beneficial. Third, while crowdsourcing is often useful for obtaining relevance judgments for Web search, it does not work as well for E-Com search due to difficulty in eliciting sufficiently fine grained relevance judgments. Finally, among the multiple feedback signals, the order rate is found to be the most robust training objective, followed by click rate, while add-to-cart ratio seems least robust, suggesting that an effective practical strategy may be to initially use click rates for training and gradually shift to using order rates as they become available. 
\end{abstract}

\copyrightyear{2017} 
\acmYear{2017} 
\setcopyright{acmcopyright}
\acmConference{SIGIR '17}{}{August 7--11, 2017, Shinjuku, Tokyo, Japan}\acmPrice{15.00}\acmDOI{http://dx.doi.org/10.1145/3077136.3080838}
\acmISBN{978-1-4503-5022-8/17/08}

\fancyhead{}

\maketitle

\section{Introduction}

E-Commerce (E-Com) search is an important emerging new application of information retrieval. Virtually all major retailers have their own product search engines, with popular engines processing millions of query requests per day. As E-shopping becomes increasingly popular, optimization of their search quality is increasingly important since an improved E-Com search engine can potentially save all users' time while increasing their satisfaction. 

Due to its importance, E-Com search has recently attracted increasing attention and has been studied from multiple perspectives~\cite{Li:2011,Duan:2013,VanGysel:2016,Vandic:2013,Long:2012,Chen:2011,Yu:2014} (See  Section~\ref{sec:related} for a more detailed review of all the major work).  However, one of the most central questions in E-Com search: How to optimally apply learning to rank (LETOR) methods for ranking of products, has attracted little attention. 

Over the past decade, Learning to Rank (LETOR) methods, which involve applying machine learning techniques on ranking problems, have proven to be very successful in optimizing search engines; specifically, they have been extensively studied in the context of Web search~\cite{chapelle2011yahoo,li2014learning,tax2015cross,azzopardi2016advances} to combine multiple features to optimize ranking. 
Thus, not surprisingly,  learning to rank is also the backbone technique for optimizing the ranking of products  in product search. 

While there has been much progress made in research on learning to rank and many LETOR methods have been proposed (see, e.g., \cite{liu2009learning,li2014learning}), applications of these methods to any search engine optimization would still face many practical challenges, notably how to define features, how to convert the search log data into effective training sets, how to obtain relevance judgments including both explicit judgments by humans and implicit judgments based on search log, and what objective functions to optimize for specific applications. Compared with the vast amount of work on improving algorithms and models for learning to rank, there has been little work on studying these practical challenges; yet they must be tackled in order to effectively deploy a LETOR method for E-Com search applications. 

In this paper, we study these practical challenges in applying LETOR to E-Com search with a focus on addressing research questions
related to 1) Effective feature representation; 2) Effectiveness of crowdsourcing relevance judgments; and 3) Exploitation of multiple feedback signals.  As no previous work has reported results on comparing the major LETOR methods for E-Com search, we first compare multiple representative major LETOR methods on an industry data set to see how well they perform for E-Com search; our results show that 
LambdaMART, one of the best performing methods for Web search, has also performed very well for E-Com search, outperforming all the other methods that we evaluated. Next, using LambdaMART as our LETOR method, we explore various issues related to feature representation, including  particularly the effectiveness of popularity-based features and the issue of query attribute sparsity. 
Our results show that popularity-based features, despite their independence of queries, are clearly effective and LambdaMART is able to optimize their combination with relevance-based features. The sparsity of query attribute features, however, poses challenges for LETOR methods, and avoiding to use sparse attributes via feature selection is found to be beneficial.

Next, we study the reliability of using crowdsourcing to obtain relevance judgments for E-Com search. Since the power of LETOR mainly comes from high-quality training examples, how to obtain reliable relevance judgements is an important question. Crowdsourcing is a commonly used approach to obtain relevance judgments in an affordable way \cite{alonso2008crowdsourcing} and has been working well for Web search \cite{le2010ensuring}, but due to the fact that the relevance criterion in E-Com search tends to be more complex than Web search (users of an E-Com search engine have very strict preferences for the products that they would eventually want to buy), it is questionable whether we can reliably obtain useful relevance judgments for E-Com search via crowdsourcing. To address this question, we analyze the quality of the relevance judgments created via crowdsourcing by leveraging the search log data. The results indeed show that the quality of crowdsourced judgments is low with a significant number of the products of the highest-level relevance with zero or just one click according to the search log data, even though they have been presented to users many times, suggesting that although crowdsourcing has been useful for obtaining relevance judgments for training LETOR methods in Web search,  it does not work as well for E-Com search. 

Finally, we study how to best exploit the different feedback signals that we discussed above for LETOR by using LambdaMart as the learning algorithm and varying the combinations of these feedback signals for training and testing. Our experiment results show that 
among all the user feedback signals, click rates appear to be the easiest to predict, while add-to-cart ratios  the hardest. Training to optimize order rates is found to be most robust (followed by click through rates) in predicting various other objectives (i.e., click rates, add-to-cart-ratios, and revenues), often delivering the best or near-the-best performance indistinguishable from training on the target signal itself. The closely related revenue, however, behaves less robustly. These findings suggest that
a reasonable simple strategy for an E-Com search engine is to use click rate based models for query segments where sufficient order data is not available, but switch to order rate based models once it is.

In summary, our paper conducts the first systematic study of challenges in applying LETOR methods to E-Com search and makes the following contributions:
\begin{enumerate}
\item We introduce and study multiple new challenges in applying LETOR to E-Com search, including relevance vs. popularity features, query attribute sparsity, difficulty in accurately eliciting human relevance judgments, complexity in optimal exploitation of multiple feedback signals.  
\item We show that popularity-based features are very effective despite their independence of queries and LETOR methods can effectively optimize their combination with other relevance-based features.
\item We show that the sparsity of query attributes poses challenges for LETOR and selecting features to avoid or alleviate sparsity is beneficial.

\item We show that while effective for Web search, crowdsourcing is not very effective for obtaining relevance judgments for E-Com search. 
\item We show that among all the feedback signals,  order rate is the most robust (followed by clickthroughs), suggesting that an effective practical strategy may be to initially use click rates for training (due to their availability) and gradually shift to using order rates as they become available.  
\end{enumerate}

\section{Learning to Rank for E-Com Search}

In this Section, we provide some background on learning to rank for E-Com search to motivate our work and provide
 a context for the research questions that we study in this paper. We first give a general introduction to LETOR and then discuss multiple practical challenges in applying LETOR to E-Com search.

\subsection{Overview of LETOR}

Optimization of ranking of products is the primary goal of an E-Com search engine. As in other retrieval system applications, traditional retrieval models, such as BM25 and language modeling approaches, play an important role in E-Com search to enable matching of the queries from users with product descriptions in the product collection. However, content matching, while very important, is not the only signal for ranking products, and it is desirable to bring in many other potentially useful signals to improve ranking. 
Specifically, an E-Com search engine would be able to naturally accumulate  large amounts of user interaction data, including particularly user queries, clickthroughs, add-to-carts, order rates, revenue information, which can all be leveraged to improve ranking. Learning to rank methods  provide a principled way to combine a large number of features optimally and have been proven very effective for Web search \cite{liu2009learning,li2014learning}. 

The basic idea of all these LETOR methods is to assume that given a query, the scoring of  items to be ranked is a parameterized function of multiple features computed based on the query, the items and any other relevant information where the parameters are generally to control the weights (contributions) of different features. These parameters can then be learned by optimizing the performance of the parameterized ranking function on a training data set where we not only have queries and items, but also relevance judgments to indicate which items should have been ranked high for a query. Once the parameters are learned, the ranking function (now with parameter values that have been learned from a training set) can be applied to any future query to rank the items with respect the query. The traditional retrieval functions such as BM25 are generally the most elementary building blocks (features) in these LETOR methods, but LETOR enables us to use many more different kinds of features.

\subsection {Application of LETOR to E-Com Search}\label{sec:challenges}

Successful application of learning to rank methods for E-Com search requires us to optimize multiple practical decisions. The first is to choose a good LETOR model. Given the same data, is it better to learn a single model across the board, or train multiple models for different segments?
How do well known learning to rank models perform for the task?
In particular, are LambdaMART models, which perform very well for Web search,  still the best?

Next there are important decisions 
 regarding training dataset creation (e.g. feature representation, source of relevance judgments etc.) and model learning (e.g. objective to be optimized, model to use etc.). In particular the trained models should be capable of generalizing to:
\begin{enumerate}
\item Previously unseen queries not in the training set
\item Previously unseen documents to be ranked for queries seen in the training set
\end{enumerate}

Note that, we use the term document/product interchangeably throughout the paper. Below we discuss the various challenges associated with achieving these modes of generalization, and highlight the relevant research questions that we will focus in our study.

\subsubsection{Feature Representation:}
Successful application of LETOR methods often depend upon the construction of useful features.  We can organize ranking features into three groups:

\begin{enumerate}
\item Query Features: These features are purely query properties. Eg. Query length, expected product category etc.
\item Document Features: These features are purely document properties. Eg. Title length, User Ratings, Total sales, department etc.
\item Query-Document Features: These features are properties of a query-document pair. Eg. BM25F text match score, Whether document belongs to the department predicted for the query etc.
\vspace{-3mm}
\end{enumerate}

\begin{figure}[!htb]
        \centering
        \includegraphics[width=85.0mm]{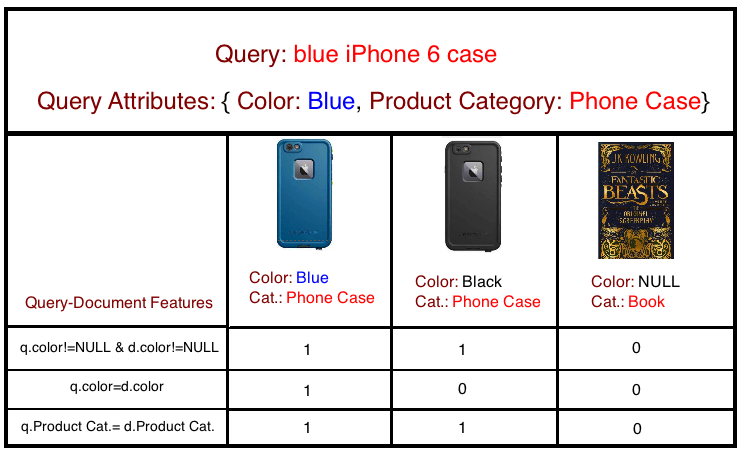}
        \vspace{-7mm}
        \caption{Features in an E-com query}
        \label{fig:f_att}
        \vspace{-2mm}
\end{figure}

A key aspect of E-Com search is the presence of a large number of product and query attributes, around which ranking features need to be constructed. Product attributes are properties of products such as Brand, Rating, Category etc. present in the catalog either manually provided by the merchant or automatically extracted from the description.
Query attributes are defined as the attributes expected by the query in the products. Macdonald et. al.  investigated which types of query attributes are useful to improve learned models~\cite{macdonald2012usefulness}. These query features are typically obtained by either parsing the query directly, analyzing search log data, or training an attribute predictor for the query. For example a query like \textit{blue iPhone 6 case} expects products with attribute ``Color: Blue'' and ``Category: Phone Cases''. We refer to these as query attributes. These are then matched with document attributes to generate query-document features. An example is shown in Figure \ref{fig:f_att}.

There are three specific challenges related to feature representation as we will discuss below. 

\noindent {\bf 1. Balancing of Relevance and Popularity:}
Figure~\ref{fig:f_att}  highlights two scenarios of challenges in optimizing the combination of features related to the balancing of relevance and popularity: 

\begin{enumerate}
\item Catalog has several ``blue'' iPhone cases. In which case the ranker should be able to rank more popular matching products higher. Popularity/ratings/sales etc. are purely document properties which now become important.

\item Catalog has several ``black'' iPhone cases which are more popular than the ``blue'' iPhone cases. Apart from popularity, the black and blue iPhone cases differ only in 1 feature, i.e., the color. Document specific features which are primarily meant to resolve among equally relevant products, should not be allowed to promote irrelevant products. 
\end{enumerate}

More importantly certain query attributes may be more critical than others. The user may be open to buying a more popular ``black'' iPhone case even on a ``blue iPhone case'' query, but certainly would not be willing to buy an iPhone. Thus while in some cases allowing popularity to dominate over color match related features may be acceptable, letting it dominate over category match is certainly unacceptable. 

Thus an interesting question here is whether LETOR methods can optimize the balancing of 
relevance and popularity and whether the popularity-based features are effective for E-Com search. The relevance vs popularity problem is especially severe in linear rankers, and we study whether a non-linear ranker such as LambdaMART can do better.

\noindent {\bf 2. Query attribute sparsity:}
While a given product will typically have $<100$ attributes associated with it, due to substantial diversity in the types of products in an E-Com catalog, one can easily expect $1000$s of unique attributes to be present in the catalog. 
The number of attributes predicted for a query tend to be even smaller. This can cause problems for unseen queries. It is quite possible that an unseen query has a set of attributes rarely observed in the training set.  To what extent does the sparsity negatively impact the performance? How can we train our models to still work well for these queries? One option is to try to gather more data on these queries, but this is generally expensive. More importantly there are always going to be unseen queries. An alternate option is to train a separate segmented ranker for such queries on a subset of overlapping features. We explore the conditions under which such a segmentation would make sense.
We compare whether its better to use the same ranker or train a different one based on the intersection of attributes.

\noindent {\bf 3. Engagement features:}
Engagement signals such as clicks, cart-adds, orders etc. are computed based on user interactions. As we discuss in the next section, they also serve as surrogates of relevance judgments. Due to this strong correlation, they tend to cause overfitting, and therefore we do not use them as features in our experiments.

\subsubsection{Relevance judgments:}

One common challenge in applying LETOR is to obtain reliable relevance judgments so that one can create high-quality training data sets for training a LETOR method. The effectiveness of LETOR clearly depends on the quality of training data.
Standardized web search test collections are based on query-document relevance ratings elicited from human experts and crowdsourcing ~\cite{DBLP:journals/corr/QinL13}. This is not a viable option in the E-Com domain, where the relevance ratings need to reflect the utility of a product to the user rather than merely text relevance or intent match with the query. Utility is typically a complex function of a product not only matching the query intent, but also the product's brand, price, quality, value for money, associated warranty, availability etc. For example for a query like \textit{14 inch laptop}, there are typically $100$s of matching laptops which are all equally relevant from a human rater's perspective. However the likelihood of a user purchasing them varies significantly from one to the other, based on brand, quality, hardware specs, operating system, reviews, battery life, price, warranty etc~\cite{alonso2009relevance}, making it challenging to make relevance judgments; indeed,  in such cases human judgments are not sufficient to provide a fine grained ground truth ordering of products for the query unless we have users to make judgements for their own queries.

In the absence of human judgements, we must turn to relevance feedback signals available from search logs for ground truth creation. In contrast with web search which only has clicks, E-Com search logs contain four prominent relevance feedback signals: clicks, cart-adds, orders and revenue. They represent potentially different training objectives as explained below. Formal equations for their computation are provided in section \ref{sec:dataset}.

\vspace{-1mm}
\begin{enumerate}[leftmargin=0cm,itemindent=.5cm,labelwidth=\itemindent,labelsep=0cm,align=left]
\item \textbf{Perceived utility of search results (Click through rate):} Function of users' perceived utility of a product, assessed based on the resulting snippet, typically title, image, price and ratings etc. Computed as the ratio of clicks a product receives for a query and its impressions (number of times shown to the user) for the query. This is closest to a typical web search setting. It especially makes sense for E-Com search engines which are relatively nascent or when churn in product or query distributions is high, so sufficient order/add-to-cart data is not available.

\item \textbf{Perceived utility of the product page (Add-to-cart ratio):} A function of users' perceived utility of a product, assessed based on the product's webpage containing comprehensive information such as detailed description, attributes, multiple images/videos and availability etc. Computed as the ratio of add-to-carts a product receives for a query and its number of clicks for the query. In case of high spam especially in online market places, this is a more suitable objective. 

\item \textbf{Overall user satisfaction (Order rate):} A function of users' overall satisfaction with a product. Computed as the ratio of orders a product receives for a query and its impressions for the query. It encompasses users' perceived utility based on both result page and item page information. Order data while being a stronger signal of user satisfaction, tends to be sparser compared to clicks and add-to-carts.

\item \textbf{Business value (Revenue rate):} Based not only on user satisfaction metrics like the previous three, but also the revenue obtained from a product's purchases. Computed as the ratio of revenue generated by a product for a query and its impressions for the query. It has the same level of sparsity as order rate.
\end{enumerate}

Choosing the right objective however is non-trivial. Clearly increasing available training data improves model quality, but high-quality signals tend to be sparse, leading to a tradeoff between the quantity and quality of training data. Understanding this trade-off is crucial to generating training datasets. To the best of our knowledge, no prior works have studied this question in E-Com domain.

One can also choose arbitrary combinations of these engagement metrics to define a target~\cite{tang2016empirical}. This is non-trivial and is often done based on several online experiments. For this work, we will focus more on the relationship between these objectives.  We will systematically examine these different types of feedback signals and experimentally study how to best use them.

\section{Experiment Design}

The main goal of our experiments is to study the new challenges in applying LETOR to E-Com search as we discussed in section~\ref{sec:challenges}. Below we will describe the data set we used,  the implementation of LETOR methods, and our experiment procedure for each of the research questions we study.

\subsection{Dataset}\label{sec:dataset}

Our E-Com data set consists of $2.8K$ randomly selected product search queries, and a catalog of $5M$ product/documents\footnote{A subset obtained from www.walmart.com product catalog. We use the term ``product"/``document" interchangeably throughout the paper.}. For each query, we retrieved top $120$ products using a $BM25F$\cite{robertson1995okapi} based retrieval algorithm. For each query-document pair $(q,d)$, we then collected statistics on impressions\footnote{Number of times a product was shown to the user for a query} $(imp(q,d))$, clicks $(clicks(q,d))$, add-to-carts $(atc(q,d))$, orders $(orders(q,d))$ and revenue $(rev(q,d))$ from search logs. Based on these statistics, we then assigned relevance ratings as follows.

\noindent For each query,  we eliminated products for which less than $100$ impressions were observed in the search log, to reduce variance in rate estimates. One can also use alternate smoothing strategies here~\cite{hardle2012smoothing}, but since these have not previously been tested on E-Com data and we had sufficient data available, we decided to simply drop the low impression products/documents. Post the filtering step, we had on average $94.2$ documents per query. Each $<$query, document$>$ pair is considered as a single training instance to the LETOR methods.\\
\noindent Let $D_q$ be the set of documents selected for a query $q$ after the first step. Then the click/order/revenue rates and add-to-cart ratios for $q$ and $d\in D_q$ were computed as follows: 

\vspace{-2mm}
\begin{eqnarray*} 
ctr(q,d)=\frac{clicks(q,d)}{imp(q,d)}&,&or(q,d)=\frac{orders(q,d)}{imp(q,d)}\\
revr(q,d)=\frac{revenue(q,d)}{imp(q,d)}&,&atcr(q,d)=\frac{atc(q,d)}{clicks(q,d)}
\end{eqnarray*}

\noindent Finally, ground truth relevance ratings based on the above objectives were computed for all products in $D_q$ by normalizing and discretizing to a $5$ point integer scale ($0-4$). For example, the formula for the ground truth relevance ratings in case of objective click rate, i.e., ctr, is presented below:
\vspace{-2mm}

\begin{equation}
rel_{ctr}(q,d)=ceil\left(4.\frac{ctr(q,d)}{\max_{d\in D_q}{ctr(q,d)}}\right)
\end{equation}

\noindent For each query, documents with the highest $ctr$ always received a rating of $4$, while documents with no clicks received $0$. Ground truth relevance ratings based on other objectives ($rel_{atcr},rel_{or},rel_{revr}$) were computed similarly. 

Note that $atcr$ unlike other rates is obtained by normalizing with $clicks$ instead of $impressions$.  Consistent with our second objective of product page utility, it represents the empirical probability of a user being satisfied by a product after the user has clicked on it and reviewed its product page.

\begin{figure}[!htb]
        \includegraphics[width=85.0mm,height=60.0mm]{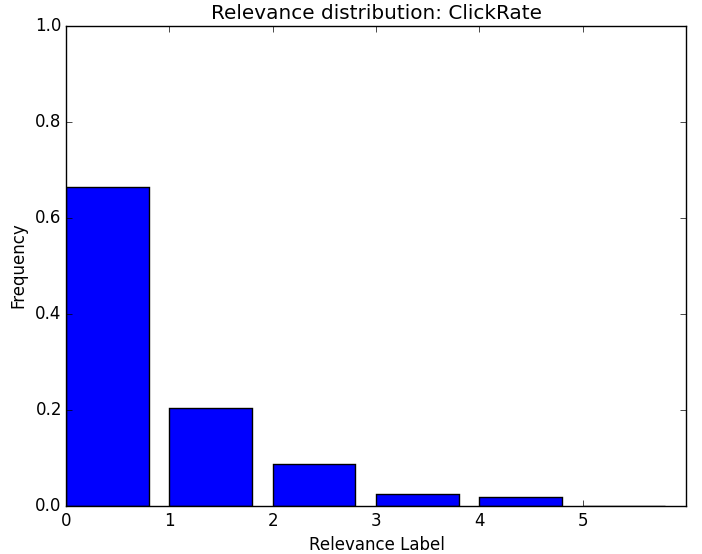}
	\caption{Relevance label distribution for the objective ``click rate''. Relevance labels are on a scale of 0-4. Average number of products per query was $94.2$. Frequencies are normalized in the scale [0,1] for each query and the average of 2.8K queries are reported in the figure.}
	\label{fig:ecom:relevance}
\vspace{-3mm}
\end{figure}

The resulting distribution of relevance labels per query for the objective ``click rate'' is shown in Figure \ref{fig:ecom:relevance}. Other relevance distributions look similar and have been omitted in the interest of space. One observes from the figure that there is a significant drop in the number of documents as we move from relevance rating $0$ to $4$. It suggests that most queries have a small number of best selling products which dominate in terms of clicks, cart-adds, orders etc. This characteristic can result from several reasons such as query specificity, competitive pricing, faster/free shipping, higher/better number of reviews and ratings, better value for money, brand loyalty etc. However, one does observe sufficient number of products associated with each relevance bucket implying that while some products tend to attract the most attention, there are indeed several products that users engage with and purchase.

Our feature set consists of $562$ features comprising various query specific, product specific and both query-product specific features. Below we highlight some of the prominent feature groups. We avoid a full list of features due to lack  of space.

\textbf{Text match}: query - product BM25F\cite{perez2010using} score, computed by matching the query with various text fields (eg. Brand, Title, Description, Department etc.) present in a semi-structured product specification document.

\textbf{Product attributes}: Overall product sales, rating, number of reviews, expected demand, brand, price etc.

\textbf{Query-Document attribute match}: Whether an attribute that takes a non-null value for the query, also takes a non-null value in the document. Similar to the first sample feature shown in Figure~\ref{fig:f_att}. 

\textbf{Query-Document attribute value match}: Whether the predicted query attribute value matches the document attribute value. Similar to the second and third sample features shown in Figure~\ref{fig:f_att}. One feature for each type of attribute (eg. Category, Brand, Price, Color, Size etc.) available in the product catalog.

Note that both attribute match and attribute value match features were real valued. The value of the feature was the same as the prediction probability of the query attribute or attribute-value.

\begin{footnotesize}
\begin{figure}[!htb]
        \centering
        \includegraphics[width=85.0mm,height=60.0mm]{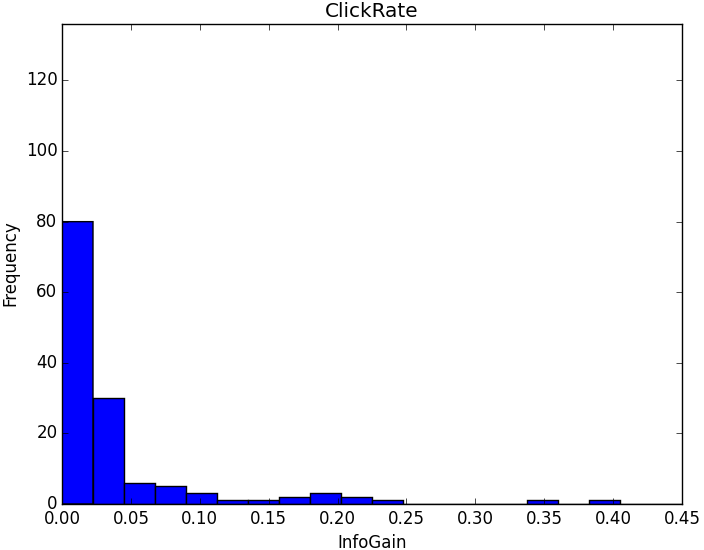}
        \vspace{-5mm}
        \caption{Feature Info-Gain Distribution}
        \label{fig:InfoGain_Distribution_ClickRate}
 \vspace{-2mm}
\end{figure}
\end{footnotesize}

To give a deeper insight into the informativeness of our features, Figure \ref{fig:InfoGain_Distribution_ClickRate} provides a histogram of information gain values w.r.t click rate based relevance rating ($rel_{ctr}$). It is evident that most of our features provide a moderate to low information gain w.r.t. the objective click rate, however, a few of them provide a high information gain w.r.t. the same objective. The relationships between these features and the other objectives are also similar and hence, omitted due to the lack of space. Note that we did not use any click, cart-add or order based features, in order to ensure our models generalize to previously unseen query-document pairs.

\subsection{Implementation of LETOR methods}

\begin{table}[!htb]
\begin{center}
 	\begin{tabular}{|l|c|}\hline
 	{\bf Algorithm} &{\bf Short form } \\\hline\hline
	RankNet~\cite{burges2005learning} & RNet\\\hline
 	RankBoost~\cite{freund2003efficient} & RBoost\\\hline
 	AdaRank~\cite{xu2007adarank}& ARank\\\hline
 	Random Forest~\cite{breiman2001random} & RF \\\hline
 	LambdaMART~\cite{burges2010ranknet} & LMART\\\hline
 	Logistic Regression (L1 regularized)~\cite{lee2006efficient,fan2008liblinear}& L1LR\\\hline
 	Logistic Regression (L2 regularized)~\cite{lin2008trust,fan2008liblinear} & L2LR\\\hline
 	SVM Classifier (L1 regularized, L2 Loss)~\cite{suykens1999least,fan2008liblinear} & L1L2SVMC\\\hline
 	SVM Classifier (L2 regularized, L1 Loss)~\cite{suykens1999least,fan2008liblinear} & L2L1SVMC\\\hline
 	SVM Regressor (L2 regularized, L2 Loss)~\cite{smola1997support,fan2008liblinear} & L2L2SVMR\\\hline	
 	SVM Regressor (L2 regularized, L1 Loss)~\cite{smola1997support,fan2008liblinear} & L2L1SVMR\\\hline
 	\end{tabular}
 \caption{Popular learning to rank algorithms}
 \label{table:algorithms}
 \end{center}
 \vspace{-5mm}
 \end{table} 

There are many LETOR methods proposed. Table \ref{table:algorithms} lists the popular LETOR approaches along with popular classification and regression methods that have also been used for ranking applications. For notational convenience, we assign abbreviations to each method which is used throughout the rest of the paper.

Our study is more focused on studying some unique new challenges in applying any LETOR method to E-Com search, thus in our experiments, we have  primarily used LambdaMART as our learning-to-rank model because it has been shown to work very well in Web search. However, since there does not yet exist any comparison of these popular LETOR methods on an E-Com data set, we also compare all the popular LETOR methods that we discussed above, mostly as a sanity check to see whether LambdaMART is also the best performing method for E-Com search. (As will be shown later, LambdaMART indeed outperforms all the other methods.) In these experiments, we used three different toolkits to experiment with learning to rank algorithms. For RankNet, Random Forest, AdaRank and RankBoost we used the RankLib toolkit~\footnote{http://www.lemurproject.org/} with default parameter settings. For LambdaMART, we used the jForests toolkit \footnote{https://github.com/yasserg/jforests}. All LambdaMART models were trained using the jforests config
properties shown in table~\ref{table:param_lambdamart}. For Logistic regression and Support Vector Machines we used the LibLinear toolkit \cite{fan2008liblinear} with default parameter settings.  In each case, we report average $NDCG@10$ \cite{jarvelin2002cumulated} computed using $5$ fold cross validation, as the performance metric.

\begin{table}[!htb]
\begin{center}
 	\begin{tabular}{|l|c|}\hline
 	{\bf parameter name} & {\bf value}\\\hline\hline
	trees.num-leaves & 7\\\hline
	trees.min-instance-percentage-per-leaf & 0.25\\\hline
	boosting.learning-rate & 0.05\\\hline
	boosting.sub-sampling & 0.3\\\hline
	trees.feature-sampling & 0.3\\\hline
	boosting.num-trees & 2000\\\hline
	learning.algorithm & LambdaMART-\\
	& RegressionTree\\\hline
	learning.evaluation-metric & NDCG\\\hline
	params.print-intermediate-valid-measurements & true\\\hline
	\end{tabular}
\caption{Parameter settings for LambdaMART}
\label{table:param_lambdamart}
 \vspace{-10mm}
\end{center}
\end{table} 

\subsection{Research Questions and Experiment Procedure}\label{sec:questions}

We will study the following research questions in this paper; for each question, we briefly describe the experiment procedure. 
 \begin{enumerate}

\item How well can LETOR methods optimize the combination of relevance-scoring features and static document-based features such as popularity features? Are popularity features useful? We can address these questions by experimenting with using only relevance-scoring features and using both relevance features and popularity features. 

\item How well can LETOR methods handle sparse query attributes? Would it be beneficial to select features to avoid sparse attributes? We can address these questions by  varying the features used and comparing using all the features (more features, but causing sparsity) and using fewer features (losing some information, but avoiding sparsity). 

\item How reliable is crowdsourcing relevance judgments for E-Com search? How is the utility of such relevance judgments compared with using naturally available user feedback signals such as clicking data, add-to-cart ratio, and order rates? We address these questions by obtaining relevance judgements through crowdsourcing and analyzing it against the user behavior signals from the activity log data. 

\item What's the best way to exploit the multiple feedback signals? Increasing available training data improves model quality, but high-quality signals tend to be sparse, leading to a trade-off between the quantity and quality of training data. How does this tradeoff impact performance? Can we train a model to optimize one objective, e.g., click rate, for which sufficient data is available easily and then apply the model to predict some different objective, e.g., order rate? We address these questions by combining different feedback signals for training and testing respectively.

\end{enumerate}

\section{Experiment Results}\label{sec:results}

In this section, we discuss our experiment results. 

\subsection{Comparison of LETOR methods}

As no previous work has compared the major LETOR methods for E-Com search, we first make such a comparison to help 
us determine the best-performing method for further study of other research questions that we discussed earlier. 

\begin{table*}[!htb]
\begin{center}
	\begin{tabular}{|l||c|c||c|c||c|c||c|c|}\hline
	& \multicolumn{2}{|c|}{\bf Click Rate} &  \multicolumn{2}{|c|}{\bf Cart Add Rate} &\multicolumn{2}{|c|}{\bf Order Rate}&\multicolumn{2}{|c|}{\bf Revenue} \\\cline{2-9}
	{\bf Algorithm} &{\bf Train } & {\bf Test}&{\bf Train}&{\bf Test}&{\bf Train}&{\bf Test} &{\bf Train}&{\bf Test} \\\hline\hline
	RNet     & 0.6857 & 0.6855  & 0.4399 & 0.4402  & 0.7158 & 0.7142 & 0.7577 & 0.7578\\\hline
	RBoost & 0.5899 & 0.5904  & 0.4073  & 0.4043  & 0.5007 & 0.4994 & 0.5663 & 0.5639\\\hline
	ARank  & 0.6877 &  0.6857 & 0.4464  & 0.4401  & 0.7334 & 0.7349 & 0.757 & 0.7566\\\hline
	RF        & 0.6378 &  0.6125 & 0.4588 & 0.4296  & 0.5707 & 0.5288 & 0.6463 & 0.5959\\\hline
	LMART & 0.8426 & {\bf 0.8291}  & 0.7664 & \textbf{0.7324}  & 0.7728 & \textbf{0.7687} & 0.8183 & \textbf{0.7998}\\\hline
	L1LR    & 0.6284 &  0.6272 & 0.4274  & 0.4252  & 0.6677 & 0.6632 & 0.6873 & 0.6822\\\hline
	L2LR    & 0.5889 & 0.5866  & 0.4066  & 0.4025  & 0.5045 & 0.4983 & 0.5751 & 0.5675\\\hline
	L1L2SVMC & 0.6366 & 0.6317  & 0.4348 & 0.4331  & 0.6870 & 0.6794 & 0.7105 & 0.7059\\\hline
	L2L1SVMC & 0.4596 &  0.4594 & 0.3274 & 0.3219  & 0.4281 & 0.4289 & 0.4503 & 0.4462\\\hline
	L2L2SVMR & 0.2358 & 0.2341  & 0.1909 & 0.1914  & 0.2100 & 0.2087 & 0.2030 & 0.2027\\\hline
	L2L1SVMR & 0.2876 &  0.2865 & 0.2110 &  0.2096 & 0.2078 & 0.2038 & 0.2093 & 0.2121\\\hline
	\end{tabular}
\caption{Comparison of ranking algorithms in terms of NDCG@10 for target variable  ``Click Rate", ``Cart Add Rate", ``Order Rate" and ``Revenue"}
\label{table:performace}
\vspace{-7mm}
\end{center}
\end{table*}

Table~\ref{table:performace} presents the summary of the results of applying different state-of-the-art LETOR algorithms on the E-com data set. We hypothesized that LambdaMART would be the best based on its superior performance in Web search, and the results indeed confirm this hypothesis; it achieves the highest test performance for each target objective, followed by AdaRank and RankNet. This observation is consistent with prior benchmark studies on web search dataset~\cite{tax2015cross}. Linear classification based approaches such as L1 regularized Logistic Regression (L1LR) and L1 Regularized L2 loss SVM Classifier (L1L2SVM) also perform well. This is explained by the fact that these methods do a good job of separating the irrelevant ($0$ rated) from relevant (1-4 rated) products, but may not be very good with correctly ordering the relevant documents among themselves. Still, since a large percentage of documents in E-com dataset are $0$ rated (Figure \ref{fig:ecom:relevance}) and our raw features are sufficiently informative, the ``simple'' linear models are able to achieve reasonable ranking performance. Finally, we observe that if a method does well on one objective, it tends to do well on others as well.

\subsection{Feature Representation}

With LambdaMART selected as the LETOR method for further experiments, we now turn to the questions about feature representation, particularly the effectiveness of popularity-based attributes as features and the  impact of query attribute sparsity, to be discussed below, respectively. 

\subsubsection{Results on popularity based attributes:}
As discussed earlier, due to the risk of the potential dominance of popularity-based features over relevance features, we cannot assume that popularity-based features are definitely effective. We thus setup an experiment to verify the usefulness of popularity based attributes in capturing the target objective. Without loss of generality, we set ``order rate", which is also the highest quality feedback available form the user activity logs, as the optimization objective. We created two different sets of training sets which vary only in the set of features that they contain. The first training set contains all the $562$ features available in the dataset, while the second one discards the popularity based features from the first training set. For each training set, we then trained a separate LambdaMART model and tested over the same testing set for both trained models. This whole process constitutes one out of the 5 folds in 5-fold cross validation, thus, we repeated this process four more times with different partitions of the data as testing data. We report the average results of the 5-fold cross validation experiment in Table~\ref{table:popularity}. This table highlights that product popularity related features which are purely document attributes are indeed useful for ranking and suggests that LambdaMART was able to learn appropriate weights to balance relevance-based and popularity-based features. Both training and testing NDCG@10 drops significantly when we remove popularity features such as total sales, reviews, ratings etc.

\begin{table}[!htb]
\begin{center}
	\begin{tabular}{|l||c|c|}\hline
	{\bf Feature Set} &{\bf Training } & {\bf Testing}\\\hline\hline
	Without Popularity &  0.71053 &  0.70415 \\\hline
	With Popularity & 0.7728 &  0.7687\\\hline
	\end{tabular}
\caption{Effect of popularity attributes in predicting the target: ``order rate"}
\label{table:popularity}
\vspace{-7mm}
\end{center}
\end{table}

\subsubsection{Results on query attribute sparsity:}
We now examine the impact of query attribute sparsity on LETOR results and test the idea of selectively using only features that help avoiding the sparsity. Specifically, for this experiment, we randomly sampled $10$ departments out of a total of $26$ in the dataset. For each department, we did the following: we first removed all queries belonging to that department and used them to create a test set. Of the remaining queries in the dataset,
we created a training set first by a) using all available attribute features and  b) using only attribute features that appear in the test set
i.e. the intersection set. We again chose ``order rate" as the target objective as this is the highest quality feedback available from the user activity logs. For each training set, we trained a separate LambdaMART model and tested it on the testing set created as described above. Table~\ref{table:sttribute_sparsity} presents the summary of these results~\footnote{Departments refer to typical departments found on a generic eCommerce site. Actual names have been removed for confidentiality.}. It clearly shows that using the intersection set leads to improved results. For example, using the intersection set of features improves LambdaMART testing NDCG@10 by almost 12\% for the queries of department ``C" , around $13.4$\% for the queries of department ``F"  etc. The number of query-document pairs available in the test and train sets are also listed. These results suggest that there is great potential to apply transfer learning and domain adaptation to LETOR for E-Com search, which should be a very interesting future direction for further study.

\begin{table}
	\begin{tabular}{|c||c|c|c|c|c|}\hline
	{\footnotesize\bf Dept.} & {\footnotesize\bf Training } & {\footnotesize\bf Testing } & {\footnotesize\bf NDCG@10} & {\footnotesize\bf NDCG@10} & {\footnotesize\bf \%}\\
	& {\footnotesize\bf $<$q,d$>$} & {\footnotesize\bf $<$q,d$>$} & {\footnotesize\bf All} & {\footnotesize\bf Intersected} & {\footnotesize\bf increase}\\
	& {\footnotesize\bf pairs} & {\footnotesize\bf pairs} & {\footnotesize\bf features} & {\footnotesize\bf Features} & {\footnotesize\bf }\\\hline
	A & 189602 & 44207 & 0.7712 & 0.8504 & 10.2570\\
	B & 233109 & 700 & 0.6940 & 0.7533 & 08.5443\\
	C & 226971 & 6838 & 0.7692 & 0.8675 & 12.7748\\
	D & 193615 & 40194 & 0.7859 & 0.8501 & 08.1733\\
	E & 210327 & 23482 & 0.7827 & 0.8732 & 11.5662\\
	F & 229258 & 4551 & 0.6986 & 0.7923 & 13.4090\\
	G & 214854 & 18955 & 0.7626 & 0.8512 & 11.6248\\
	H & 222374 & 11435 & 0.8236 & 0.8908 & 08.1598\\
	I & 225416 & 8393 & 0.8434 & 0.9496 & 12.5904\\
	J & 222300 & 11509 & 0.7903 & 0.8896 & 12.5765\\\hline
	\end{tabular}
\caption{Results on Attribute Sparsity}
\label{table:sttribute_sparsity}
\vspace{-8mm}
\end{table}

\subsection{Results on crowdsourcing}

In this experiment, we study the reliability of the relevance judgements provided by the crowd workers for e-commerce queries.  We first randomly selected a query. For this query, workers were shown a product image, title and price along with the query. Different products were shown to the workers for the same query. They were then asked to rate the relevance of the product on a 0-4 rating scale. There were well defined guidelines on what each level means with 4 being ideal and 0 being non-relevant. 

After the workers have provided their relevance judgements; for each query, we filtered out the products that were rated as 4, i.e., ideal, by the workers. Next, for all these products that were marked as ideal by the workers, we looked at the user activity log to find out their actual number of clicks and impressions for that query. What we found out is that even though the crowd-workers rated all these products with the ideal rating, i.e., $4$, the actual number of clicks the products received varied widely given that each item had a sufficiently large number of impressions in top 2 ranks. 

Table~\ref{table:crowdsourcing} shows our findings for five example queries, i.e., ``Neutrogena", ``Vacuum Cleaner", ``Monster High Dolls", ``Water Bottles" and ``Outdoor Furnitures". For each query, Table~\ref{table:crowdsourcing} bins products with respect to the actual number of clicks they received. For example, the first row of Table~\ref{table:crowdsourcing} shows that $44$ items that were rated as ideal by crowd-workers with respect to query ``Neutrogena" received no clicks from real users. To easily distinguish the wide variance in the number of clicks that these products received, we report the results for the two bins, i.e., 1) products with zero clicks and 2) products with five or more clicks. To demonstrate that all these items received sufficient impressions, we show the mean, median and maximum of the number of impressions (only in top 2 ranks) that the products in each bucket received. This indeed shows that, even if the crowd-workers judged all these items to be ideal with respect to the query, user click counts suggests the actual utility of the product varied significantly across different products. Thus we conclude that crowdsourcing fails to provide reliable relevance judgements for E-commerce queries. This also supports the findings of the study by Alonso et. al.~\cite{alonso2009relevance} where they suggested that when users are interested in buying products, they apply different criteria when deciding relevance. This is often hard to judge by the crowd workers.

\begin{table}[!tb]\small
\begin{center}
	\begin{tabular}{|c||c|c|c|c|c|}\hline
	{\bf query} & {\bf No of} & {\bf No of}& \multicolumn{3}{|c|}{ \bf Impressions (top 2 ranks)} \\\cline{4-6}
	& {\bf clicks}& {\bf items} & {\bf mean} & {\bf median} & {\bf max}\\\hline\hline
	Neutrogena& 0 & 44 & 165.11 & 104.5 & 982\\\cline{2-6}
        & 5 or more & 16 & 4866.56 & 999.5 & 21901\\\hline\hline
        
	Vacuum& 0 & 28 & 84.46 & 56.5 & 277\\\cline{2-6}
        Cleaner& 5 or more & 13 & 3040.76 & 620 & 17799\\\hline\hline
        
        Monster& 0 & 13 & 292.61 & 231 & 759\\\cline{2-6}
        High Dolls& 5 or more & 24 & 3492.66 & 1806.5 & 30460\\\hline\hline

        Water & 0 & 8 & 187.5 & 38.0 & 884\\\cline{2-6}
        Bottles & 5 or more & 13 & 6107.07 & 1580 & 44044\\\hline\hline

        Outdoor & 0 & 8 & 107.37 & 120.5 & 229\\\cline{2-6}
        Furnitures& 5 or more & 8 & 3254.125 & 841.0 & 18733\\\hline

	\end{tabular}
\caption{Variation in relevance judgments obtained from click data for products rated as 4 (ideal) by the crowd.}
\label{table:crowdsourcing}
\vspace{-5mm}
\end{center}
\end{table}

\subsection{Cross Objective Learning Experiments}

According to~\cite{Yilmaz:2010}, the target metric
is not always the best (most informative) metric for training a LETOR method. We thus would like to experimentally understand how to best exploit the multiple training objectives available in E-Com search.
To achieve this goal, we evaluated the performance of LambdaMART models trained using one training objective on test datasets based on other objectives. The results are shown in Table \ref{table:transferLearning}, where each row corresponds to training with a different objective (e.g., the first row with numerical values is training on click rate), and each column corresponds to testing with a different objective (e.g., the first column with numerical values is testing with click rate as the objective). 

We observe that the best performance on a given test objective is indeed not always achieved by models trained on the same objective. Specifically, Table \ref{table:suitableTargets} highlights optimal training objectives that lead to statistically significantly better models for a given test objective, compared to the sub-optimal objectives. For example, both $ctr$ and $or$ based models were found to perform significantly better in predicting click rates than $atcr$ and $revr$ based models. Also in each case, we did not observe any statistically significant improvements when comparing the optimal objectives among themselves. 

Overall, for each test objective, order rate $or$ turned out to be consistently optimal, while $atcr$ turns out to be consistently sub-optimal. This highlights the importance of using order rate as a robust training objective. In general for any given query segment, one can initially build models trained on $ctr$ (which is optimal for objectives $or,ctr$ and is available easily in a significant amount) since order rates are pretty sparse on initial stages and then switch to $or$ based models once sufficient order data is available.

\begin{table*}[!htb]

\parbox{.49\linewidth}{
\center
	\begin{tabular}{|l||p{1cm}|p{1cm}|p{1cm}|p{1cm}|}\hline
	& \multicolumn{4}{|c|}{\bf Test data objective}\\\hline
	{\bf Training data } &{\bf Click }&{\bf Order }&{\bf Reve- nue }&{\bf Cart Add }\\
	{\bf objective} &{\bf rate } &{\bf rate }&{\bf rate }&{\bf ratio }\\\hline\hline
	click rate & 0.8291 & 0.7680 & 0.7943 & 0.7175 \\\hline
	Order Rate & 0.8258 & 0.7687 & 0.8006 & 0.7252 \\\hline
	Revenue & 0.8206 & 0.7694 & 0.7998 & 0.7186 \\\hline
	Cart Add Ratio & 0.8061 & 0.7564 & 0.7701 & 0.7234 \\\hline
	\end{tabular}
\caption{ Results for Cross Target Learning}
\label{table:transferLearning}
}
\hfill
\parbox{.49\linewidth}{
\center
	\begin{tabular}{|l||p{2.5cm}|p{2.5cm}|}\hline
	{\bf Test } &{\bf Optimal }&{\bf Sub-optimal }\\
	{\bf objective} &{\bf training } &{\bf training }\\
	{\bf objective} &{\bf  objectives} &{\bf  objectives}\\\hline\hline
	Click Rate & $ctr,or$ & $atcr, revr$ \\\hline
	Order Rate & $or,revr,ctr$ & $atcr$ \\\hline
	Revenue &   $revr,or$ & $ctr,atcr$ \\\hline
	Cart Add Ratio & $actr,or$ & $ctr,revr$ \\\hline
	\end{tabular}
\caption{Suitable training objectives based on stat. significant improvements (Wilcoxon's signed rank test \cite{wilcoxon1945individual} at level $0.05$)}
\label{table:suitableTargets}
}
\vspace{-5mm}
\end{table*}

\section{Related Work}\label{sec:related}

E-Com search has recently attracted increasing attention and has been studied from multiple perspectives, including e.g., economic theory-based model~\cite{Li:2011}, adaptation of traditional retrieval models for product search \cite{Duan:2013}, applying representation learning to product search~\cite{VanGysel:2016}, facet selection for product search~\cite{Vandic:2013}, beyond relevance ranking~\cite{Long:2012} and diversification of product search results~\cite{Chen:2011,Yu:2014}.

Several methods have been proposed to solve the Learning to Rank problem for web search. The best-performing methods include RankNet, LambdaMART, AdaRank and RankBoost etc.~\cite{li2014learning,liu2009learning}. Researchers have conducted comprehensive studies to compare the performance of popular LETOR approaches on benchmark web data-sets~\cite{chapelle2011yahoo,tax2015cross}. In particular, LambdaMART~\cite{chapelle2011yahoo,tax2015cross} and RankNet~\cite{tax2015cross} are known to perform particularly well. In our work, we also compared many of these methods and confirmed that LambdaMART is very effective, outperforming all the other methods that we compared with. 

Macdonald et. al.~\cite{macdonald2013whens} proposed general guidelines on best practices for deploying LETOR methods. Specifically, they provided three important general observations critical to the deployment of LETOR models. First, the smallest effective sample for LETOR varies according to several factors: the information need, the evaluation measure used to test the models, and the presence of anchor text in the document representation used for sampling. Second, the choice of the learning evaluation measure can indeed have an impact upon the effectiveness of the resulting learned model. Third, the importance of different classes of features within a learned model are dependent on both the sample size and the rank cutoff of the learning evaluation measure.  Our work extends this line of contributions by adding additional findings related to unique challenges in using LETOR for E-Com search, especially the effectiveness of popularity-based features, the effectiveness of selecting features to avoid query attribute sparsity, and findings about how to best exploit the multiple feedback signals.  In other related works, Chapelle et. al. ~\cite{chapelle2011future} lay out open challenges in Learning to rank for web search. Some prominent open challenges they mentioned include sample selection bias in training, efficiency and scalability of large scale learning to rank,  transfer learning to rank, online learning to rank etc. Our results on query attribute sparsity clearly suggest the importance of some of these techniques, particularly transfer learning to rank. 

Our exploration of crowdsourcing is related to the previous work on this topic \cite{alonso2009relevance} which discussed the different relevance criteria in E-Com, which may provide an explanation of our finding that crowdsourcing is not reliable for E-Com search.

\section{Conclusions and Future Work}

Learning to rank for E-Commerce (E-Com) search is an emerging important problem whose study is still in its infancy with many interesting new challenges to be solved. This paper presents a systematic experimental study of the application of state of the art LETOR methods to E-Com search using an industry data set. Below we summarize our experimental findings and discuss multiple interesting future research directions related to LETOR for E-Com search motivated by our study and observations.

First, the relative performance of different learning to rank methods on our product search data set is mostly consistent with past observations on web data sets, confirming that LambdaMART is  the best performing method, and can thus be recommended for use in E-Com application systems and as a reasonable baseline for further research on learning to rank for E-Com search.  

Second, the popularity-based features are very effective; despite the potential risk of imbalanced weighting of popularity-based features and relevance-based features, LETOR methods can learn to balance this weighting appropriately. However, query attribute sparsity raises challenges for LambdaMART and  selectively using a subset of  features to avoid sparsity is beneficial. 

Third, while useful for annotating Web search data sets, crowdsourcing is shown to generate mostly unreliable judgments for E-Com search. 

Finally, when using LambdaMART, among all the user feedback signals, click rates appear to be the easiest to predict, while the add-to-cart ratios  the hardest. Training to optimize order rates is found to be most robust (followed by click through rates) in predicting various other objectives (i.e., click rates, add-to-cart-ratios, and revenues), often delivering the best or near-the-best performance indistinguishable from training on the target signal itself. The closely related revenue, however, behaves less robustly. These findings suggest that
a reasonable simple strategy for an E-Com search engine is to use click rate based models for query segments where sufficient order data is not available, but switch to order rate based models once it is.

As in the case of all experimental studies of retrieval systems, it is important to use more test collections to further verify our findings, which would be an important future work. It is also very interesting to further study what is the best way to use different feedback signals, a unique challenge specific to product search (not present in Web search). For example, given that training on the same signal as the target of prediction may not be as good as training on another correlated signal, can we intelligently combine various feedback signals to obtain a more robust effective ranker? How can we optimally combine them? There are clearly multiple possibilities, such as combining them as features vs. using each to train a different model and then combining multiple models. It would really be interesting to develop a method that can switch from click-rate to order-rate, or even better, a model that can combine both and adapt based on the amount of data that is available. 

Another interesting direction is the following: since a lot of queries in E-com domain are category-related queries, e.g., show me ``tv sets'' or ``sofas'', this hints that there is a general tendency of exploratory-like search behavior in E-com. With this in mind, it would make sense to separate/classify queries into different classes, e.g., ``product'' and ``category''. Each class may require a different model and/or different evaluation measures. nDCG may be good for ``focused'' queries because these are more precision-oriented, but recall (or other measures) may be better for more abstract queries, like ``tv sets'' or ``sofas''. Finally, the unique challenges we identified in Section 2 can serve as a roadmap to help us identify potentially many interesting new research problems about using LETOR for E-Com search for future research. In particular, exploration of adaptive cutoff in NDCG that is sensitive to a query would be interesting given the high variance in browsing depth in E-Com search.

Our work has also revealed a number of interesting novel challenges for future research on LETOR in general which we briefly discuss below:

(1) Presence of Uncertain (engagement related) features: 
sales/ ratings/ reviews etc. related features are heavily dependent upon the age of the product in the catalog and 
the number of impressions a product has received. Products newly added in the catalog, typically have zero values associated with these features, and consequently get artificially penalized by the ranker. In future we intend to explore learning to rank methods capable of handling such uncertainties in feature values. 

(2) Correlated features: clicks, cart-adds and orders are clearly understood to be closely related and hence ignored from the feature set. Hidden relationships may however still exist between these engagement signals and other features. Especially in context of industry setting where several engineers may be generating $1000$s of features and the data scientists training the models may not know how exactly each was generated.   Existence of such hidden relationships can lead to artificially high performance on the train/test set being used. Thus their detection is an important challenge.

(3) Data quality issues: 
Since the data is often entered by vendors or inferred automatically, there are always issues with data quality in terms of missing attributes or incorrect attributes etc.  

(4) Query sparsity: Transfer learning to rank problems become particularly important when the goal is to generalize to unseen queries and solve some of the attribute sparsity challenges mentioned in the paper. Specially, having thousands of attribute values in products poses challenges in learning. How can we reduce the dimensionality and how much data do we need for this, is an interesting open problem. Ideas from deep neural networks may be useful here. 

(5) Query-specific variable cutoff for NDCG measure  to accommodate variable browsing depth of users: It is worth elaborating this point further. Typical web search studies have reported NDCG@10. We have very different K requirements for different queries in product search since there tend to be high variances in the browsing depth. Indeed, generic product queries such as \textit{tv, desk, office chair} etc. constitute a substantial fraction of product search traffic. Such queries usually represent exploratory intents with $1000$s of relevant products in the catalog. Depending upon the expenditure involved, a user may want to review several products, well beyond the top $10$, before making a purchase decision. From a search quality perspective, this requires the search engine to maintain a high $NDCG@K$ for $K>>10$. Thus while in Web search, people generally only pay attention to the first page of results, users have much higher variances in their browsing depth in E-Com search, making it questionable whether optimizing a metric such as NDCG at a fixed cutoff of $10$ remains an appropriate strategy. We hypothesize that it might be beneficial to use query-specific variable cutoffs when computing NDCG (which affects the objective functions used for training LETOR methods), which would be a very interesting future research direction.

\section{Acknowledgments}
We thank WalmartLabs search team, particularly Esteban Arcaute and Pranam Kolari, for providing the necessary funding and support for this work. We also thank the anonymous reviewers for their useful comments, which helped improving the quality of the paper significantly.

\bibliographystyle{ACM-Reference-Format}
\bibliography{sigproc,sigproc_new}

\end{document}